\documentclass{elsarticle}

\usepackage{graphicx}
\usepackage{hyperref}
\usepackage[group-separator={,}]{siunitx}
\usepackage{tabularx,ragged2e,booktabs,caption}

\biboptions{sort&compress}

\begin{document}

    \begin{frontmatter}

        \title{Collaborative similarity analysis of multilayer developer-project bipartite network}
		
		\author[ustc]{Deng-Cheng Yan}
		\author[ustc,wzu,myu]{Bing-Hong Wang \corref{cor}}
		\ead{bhwang@ustc.edu.cn}
		
		\address[ustc]{Department of Modern Physics, University of Science and Technology of China, Hefei 230026, P. R. China.}
		\address[wzu]{College of Physics and Electronic Information Engineering, Wenzhou University, Wenzhou 325035, P. R. China.}
		\address[myu]{School of Science, Southwest University of Science and Technology, Mianyang 621010, P. R. China.}

		\cortext[cor]{Corresponding author at: Department of Modern Physics, University of Science and Technology of China, Hefei 230026, P. R. China.}

        \begin{abstract}
            To understand the multiple relations between developers and projects on GitHub as a whole, we model them as a multilayer bipartite network and analyze the degree distributions, the nearest neighbors' degree distributions and their correlations with degree, and the collaborative similarity distributions and their correlations with degree. Our results show that all degree distributions have a power-law form, especially, the degree distribution of projects in watching layer has double power-law form. Negative correlations between nearest neighbors' degree and degree for both developers and projects are observed in both layers, exhibiting a disassortative mixing pattern. The collaborative similarity of both developers and projects negatively correlates with degree in watching layer, while a positive correlations is observed for developers in forking layer and no obvious correlation is observed for projects in forking layer.

        \end{abstract}


    \end{frontmatter}

    \section{Introduction}
    
        The last few decades have witnessed the rapid development and adoption of information technology in a variety of industries. Software, one crucial constitute of information technology, drives the innovation of society and significantly improves our daily life. Methodology of modern software engineering encourages software developers to be social and open-source minded. GitHub, an outstanding open-source and social collaborative coding platform, was launched on April 10th, 2008. It provides abundant social functionalities such as watching and forking for developers to interact with projects efficiently. Watching is a notification mechanism to inform a developer of any new pull requests and issues of a project he/she has watched. Forking makes it possible for developers to copy projects of others as their own and keep working on a new branch. More information about the social functionalities on GitHub can be found from GitHub Help \footnote{https://help.github.com/articles/watching-repositories/}.

        From the perspective of networks, the relations established by these social functionalities between developers and projects can be naturally modeled by separate bipartite networks. In fact, many systems have been modeled as bipartite networks, such as the metabolic network \cite{metabolicnetworks}, the human sexual network \cite{humansexual} and the collaboration network \cite{collaborationnetworks} and various measurements have been proposed \cite{bipartiteclusteringcoefficient,collaborativesimilarity}. In addition, great efforts have been made to characterize \cite{PhysRevE.72.046105,bipartite-correlation}, project \cite{PhysRevE.76.046115,Chinese-railway-network,one-mode-projection-multiplex-bipartite} and model \cite{PhysRevE.70.036106,Zhang20136100,PhysRevE.72.036120} bipartite networks. From a systematic view, different relations with projects reflect different aspects of developers' behaviors and treat them as a whole may provide us a panorama. In this situation, multilayer network, also addressed as multiplex network, is a suitable tool, which has recently attracted increasing attentions \cite{weighted-multiplex-networks,triadic-relations-multiplex-networks,multiplex-pageRank,modeling-correlations-multiplex-networks,k-core-multiplex-networks,information-transport-multiplex-networks}.

	    In this paper, we model the watching and forking relations between developers and projects on GitHub as a multilayer bipartite network and apply the collaborative similarity to investigate the diversity of interests. The rest of this paper is organized as follows: In Section 2, we describe the dataset preparation and the basic statistical characteristics. Then we report the empirical analysis results, including the degree distributions, the nearest neighbors' degree distributions and the collaborative similarity distributions in Section 3. We summarize our work in Section 4 and give out a brief discussion.

	\section{Datasets preprocessing and description}
	    The dataset used in this article is provided by GHTorrent Project \cite{Gousi13}, which is, as described on GHTorrent Project website \footnote{\href{http://ghtorrent.org/}{http://ghtorrent.org/}}, a scalable, queriable and offline mirror of data offered through the GitHub REST API. It  monitors the GitHub public event time line, collects the contents and dependencies of each event and store the raw data in MongoDB database. The datasets provided by GHTorrent Project come in both MySQL dump format and MongoDB dump format. Meanwhile it also provides services for accessing the dataset programmatically or through a web interface \footnote{\href{http://ghtorrent.org/services.html}{http://ghtorrent.org/services.html}}. We download the dataset in MySQL dump format dated by January 4th, 2015 \footnote{\href{https://ghtstorage.blob.core.windows.net/downloads/mysql-2015-01-04.sql.gz}{https://ghtstorage.blob.core.windows.net/downloads/mysql-2015-01-04.sql.gz}} and restore it to our MySQL database. The detailed information about the dataset in MySQL dump format is described online \footnote{\href{http://ghtorrent.org/relational.html}{http://ghtorrent.org/relational.html}} and in Ref \cite{Gousi13}.

\begin{figure}[htbp]
			\centering
		    \includegraphics[width=1.0\textwidth]{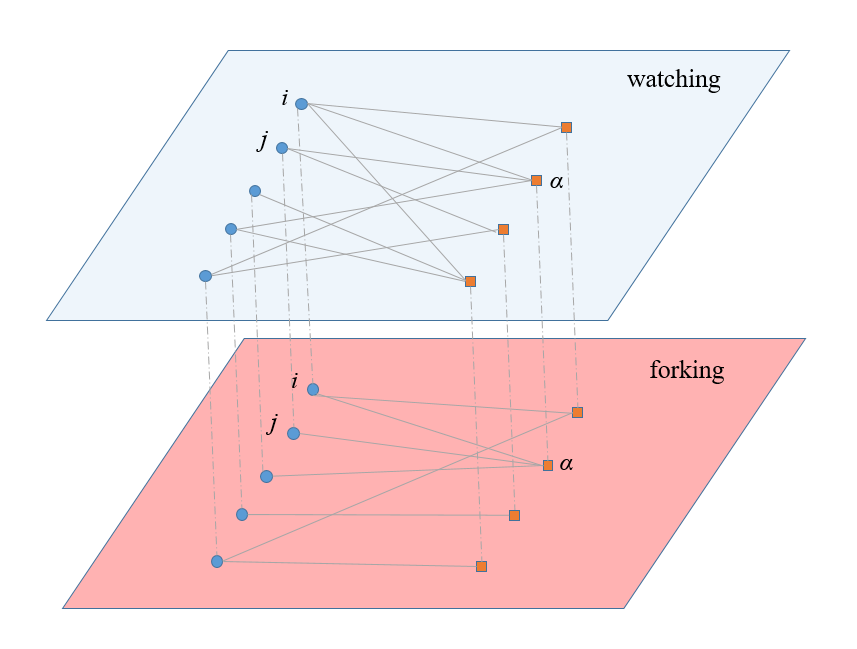}
			\caption{Sample illustration of the multilayer bipartite network discussed in this paper. The blue circles and orange rectangles represent developers and projects, respectively. There are two layers: watching layer (the light blue layer) and forking layer (light red layer). Either layer is a bipartite network representing one relation between developers and projects. For example, developer $i$ watches (forks) project $\alpha$, then there is a link (the gray line) between them in the watching (forking) layer.  The dashed line is not a real link and just connects the same node in different layers.}
			\label{fig:network-sample}
		\end{figure}

        We select the watching and forking data and extract the records that related to PHP projects. Figure \ref{fig:network-sample} is a sample illustration of the multilayer bipartite network discussed in this paper. Table \ref{tbl:dataset-statistics} summarizes the basic statistical properties of the datasets.

	    \begin{table}
            \centering
            \caption{The statistics of the datasets. $N_{d}$ and $N_{p}$ represent the number of developers and projects, respectively. $E$ denotes the number of edges. $\langle k_{d} \rangle$ and $\langle k_{p} \rangle$ denote the average degree of developers and projects, respectively.}
            \label{tbl:dataset-statistics}
            \begin{tabular}{cccccc}
                \hline\hline
                \textbf{Layer} & \textbf{$N_{d}$} & \textbf{$N_{p}$} & \textbf{$E$} & \textbf{$\langle k_{d} \rangle$} & \textbf{$\langle k_{p} \rangle$} \\
                \hline
                \textbf{watching} & 356619 & 180581 & 1094645 & 3.07 & 6.06 \\
                \textbf{forking}  & 356619 & 180581 &  461124 & 1.29 & 2.55 \\
                \hline\hline
            \end{tabular}
        \end{table}

	\section{Empirical results}
	
	    \subsection{Degree distributions}
		    \begin{figure}[htbp]
			    \centering
		        \includegraphics[width=1.0\textwidth]{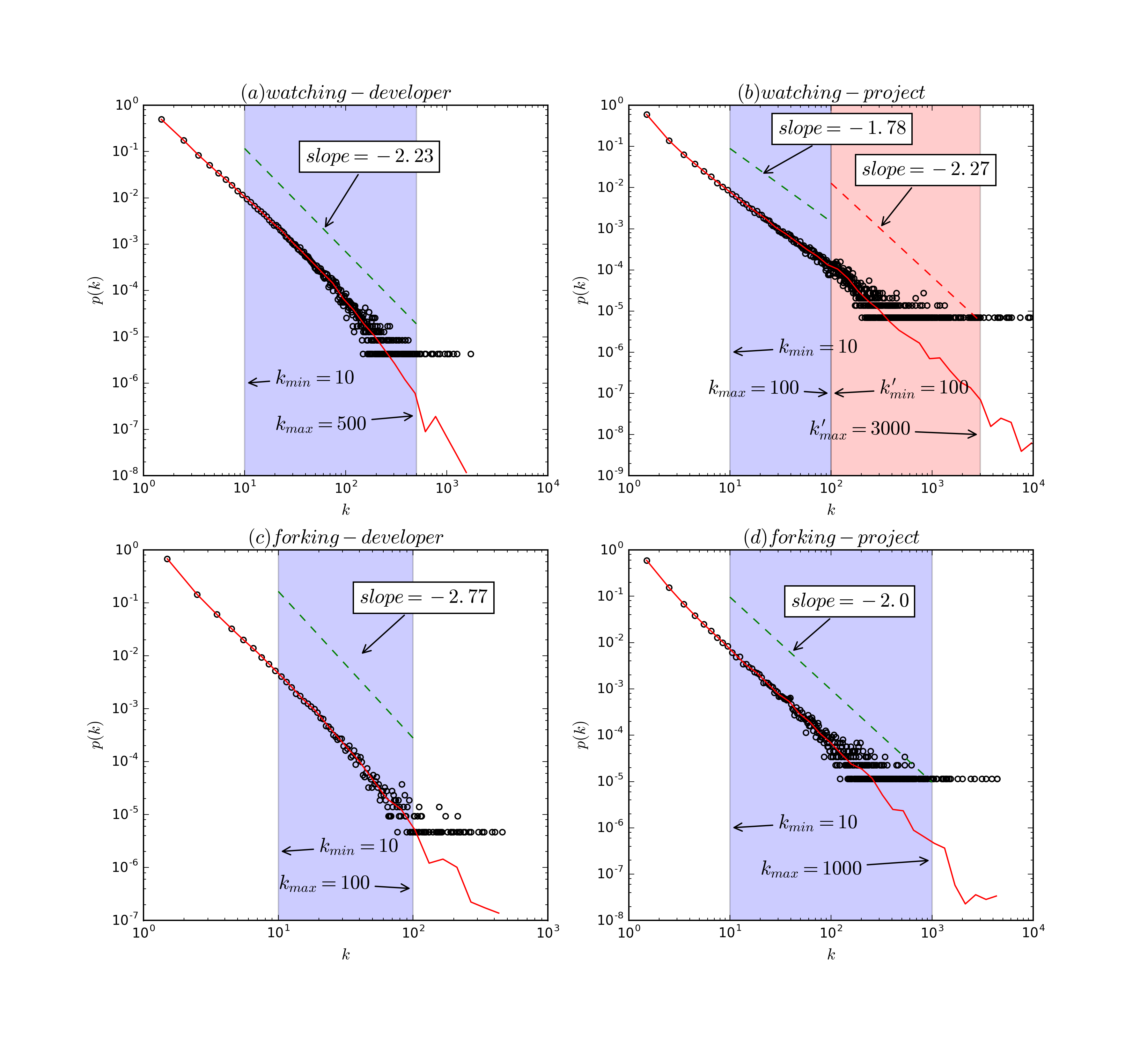}
			    \caption{The degree distributions of developers and projects in bipartite networks from watching layer and forking layer. The horizontal coordinate $k$ denotes the degree, and the longitudinal coordinate $p(k)$ denotes the probability density function of degree. Part of each distribution is fitted to power-law using powerlaw Python package from Ref. \cite{alstott2014powerlaw}. The data is treated as discrete. On log-log axes, using logarithmically spaced bins is necessary to accurately represent data (red line). Linearly spaced bins (black circle) obscure the tail of the distribution. We fix both values of $k_{min}$ and $k_{max}$. The Kolmogorov-Smirnov distances between the fitted portions of data and the fits are all less than 0.05.}
			    \label{fig:bipartite-degree-distribution}
		    \end{figure}

		    For the multilayer bipartite network, the degree of developer $i$ in layer $m$ ($m \in \{watching, forking\}$), denoted by $k_{i}^{m}$, is defined as the number of projects developer $i$ connects to in layer $m$. Similarly, the degree of project $\alpha$ in layer $m$ ($m \in \{watching, forking\}$), denoted by $k_{\alpha}^{m}$, is defined as the number of developers which have links to project $\alpha$. For example, as shown in the sample illustration of the multilayer bipartite network in Figure \ref{fig:network-sample}, $k_{i}^{watching}=3$ and $k_{\alpha}^{forking}=3$.

		    The degree distributions of developers and projects from watching layer and forking layer are reported in Figure \ref{fig:bipartite-degree-distribution} and each distribution has a heavy tail which fits the power-law form $p(k) \sim k^{-\phi}$ well. Especially, the degree distribution of projects in $watching$ layer follows a double power law.

	    \subsection{Nearest neighbors' degree distributions}
	        
		    \begin{figure}[htbp]
			    \centering
		        \includegraphics[width=1.0\textwidth]{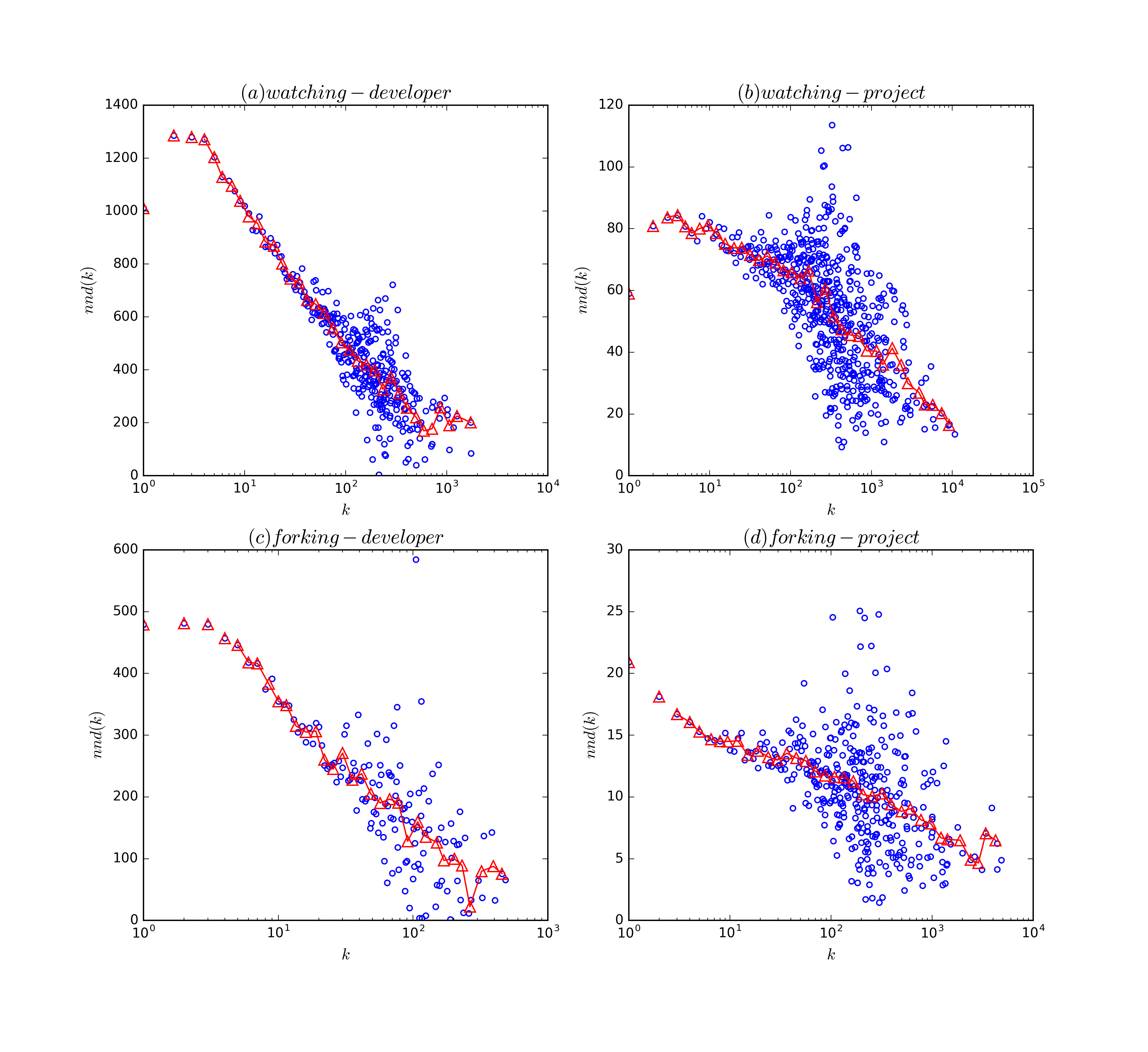}
			    \caption{The nearest neighbors' degree distributions. The horizontal and longitudinal coordinates $k$ and $nnd(k)$ represent degree and nearest neighbors' degree, respectively. The blue dots represent degree-dependent nearest neighbors' degree distributions. The degree-dependent nearest neighbors' degree fluctuates in the high-degree range due to the sparsity of data. In order to make the overall trends more clear, we apply log-binning method on the degree-dependent nearest neighbors' degree distributions. Degrees are log-binned into 40 groups and the average degree and nearest neighbors' degree of all blue dots falling in the same bin are calculated and marked as red triangles in the figures.}
			    \label{fig:nearest-neighbors-degree-distribution}
		    \end{figure}

		    The nearest neighbors' degree of developer $i$ in layer $m$ ($m \in \{watching, forking\}$), denoted by $k_{nn,i}^{m}$ is defined as the average degree over all projects developer $i$ connects to. The nearest neighbors' degree of project $\alpha$ in layer $m$ ($m \in \{watching, forking\}$), denoted by $k_{nn,\alpha}^{m}$ is defined as the average degree over all developers which have links to project $\alpha$. For example, as shown in the sample illustration of the multilayer bipartite network in Figure \ref{fig:network-sample}, $k_{nn,i}^{watching} = \frac{2+3+3}{3} = \frac{8}{3}$ and $k_{nn,\alpha}^{forking} = \frac{2+1+1}{3} = \frac{4}{3}$. The degree-dependent nearest neighbors' degree of developers in layer $m$ ($m \in \{watching, forking\}$), denoted by $nnd_{d}^{m}(k)$, is defined as the average nearest neighbors' degree of all developers of degree $k$ in layer $m$. The degree-dependent nearest neighbors' degree of projects in layer $m$ ($m \in \{watching, forking\}$), denoted by $nnd_{p}^{m}(k)$, is defined as the average nearest neighbors' degree of all projects of degree $k$ in layer $m$. For example, as shown in the sample illustration of the multilayer bipartite network in Figure \ref{fig:network-sample}, $nnd_{p}^{watching}(2) = \frac{9}{4}$ and $nnd_{d}^{forking}(1) = \frac{7}{3}$.

		    As can be seen from Figure \ref{fig:nearest-neighbors-degree-distribution}, the degree-dependent nearest neighbors' degree is negatively correlated with the degree, exhibiting a disassortative mixing pattern. This indicates that fresh developers tend to watch or fork popular projects and unpopular projects are often watched or forked by active developers, which agrees with the case of $Audioscrobbler$ and $Del.icio.us$ discussed by Shang $et al.$ \cite{collaborativesimilarity}. We improve the visualizing methods $et al.$ \cite{collaborativesimilarity} by log-binning the degree-dependent nearest neighbors' degree distributions and see more clear overall trends ranging from small-degree to high-degree than that used in \cite{collaborativesimilarity}. The degree-dependent nearest neighbors' degree distributions of projects in both layers fluctuates heavily than those of developers, indicating that the correlations of developers are stronger than those of projects, which may be explained by the difference of selection mechanisms between developers and projects\cite{collaborativesimilarity}.

	    \subsection{Collaborative similarity distributions}

		    \begin{figure}[htbp]
			    \centering
		        \includegraphics[width=1.0\textwidth]{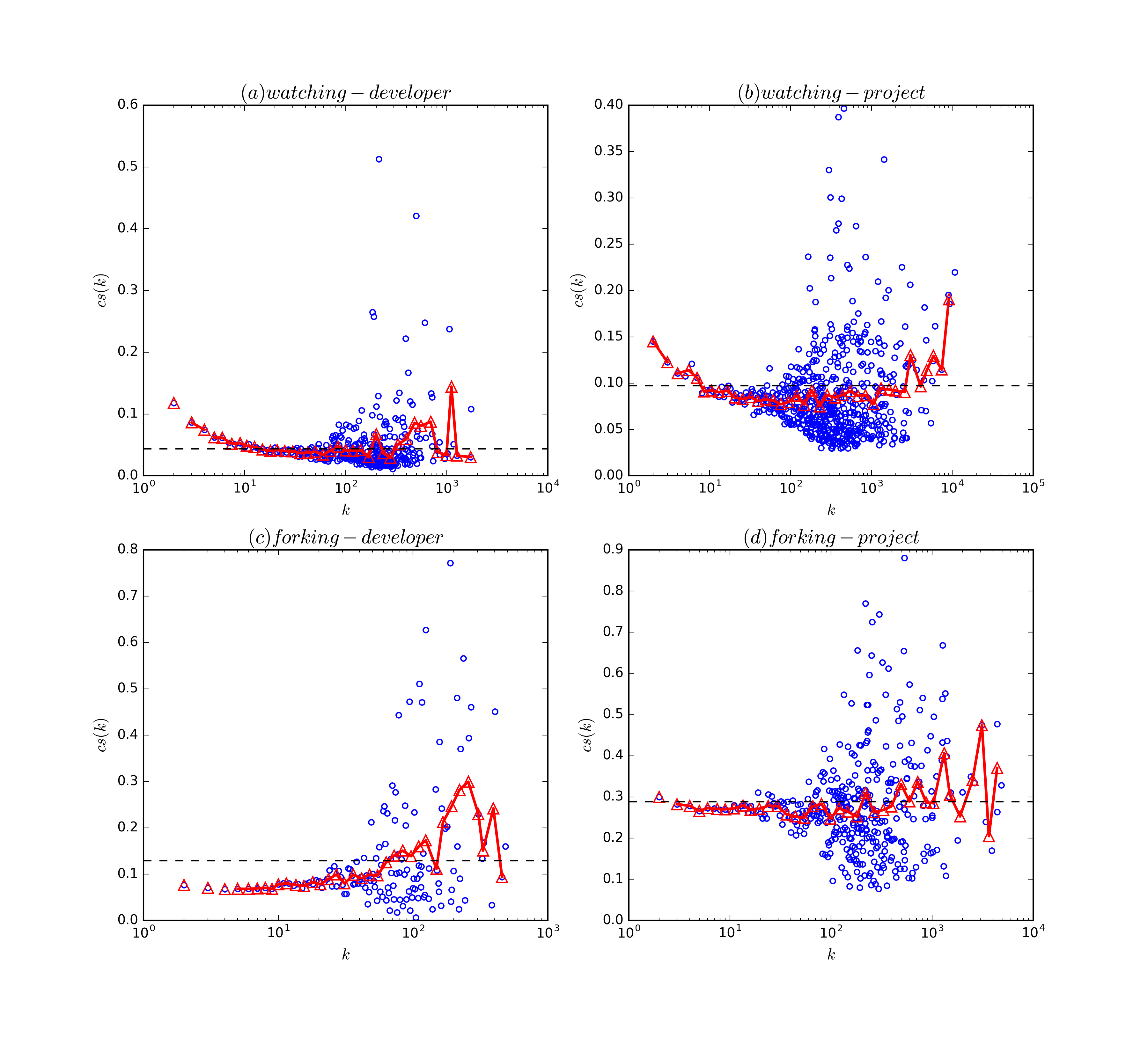}
			    \caption{The collaborative similarity distributions. The horizontal and longitudinal coordinates $k$ and $cs(k)$ represent degree and collaborative similarity, respectively. The blue dots represent degree-dependent collaborative similarity distributions. The degree-dependent collaborative similarity fluctuates in the high-degree range due to the sparsity of data. In order to make the overall trends more clear, we apply log-binning method on the degree-dependent collaborative similarity distributions. Degrees are log-binned into 40 groups and the average degree and collaborative similarity of all blue dots falling in the same bin are calculated and marked as red triangles in the figures. The dashed black lines in each figure show the overall average collaborative similarity.}
			    \label{fig:collaborative-similarity-distribution}
		    \end{figure}

		    Shang $et al.$ \cite{collaborativesimilarity} proposed a index called collaborative similarity to characterize the clustering selections from the perspective of collaborative interests instead of traditional clustering coefficient for general bipartite networks which measures the quotient between the number of squares observed and the total number of possible squares \cite{bipartiteclusteringcoefficient}. The collaborative similarity of developer $i$ in layer $m$ ($m \in \{watching, forking\}$) is defined as the average $Jaccard similarity$ between the projects which developer $i$ has selected in layer $m$ ($m \in \{watching, forking\}$) \cite{collaborativesimilarity}. The degree-dependent collaborative similarity of developers is defined as the average collaborative similarity over all developers of the same degree. Corresponding definitions for projects are similar and thus omitted here.

		    The collaborative similarity distributions are shown in Figure \ref{fig:collaborative-similarity-distribution}. We improve the visualizing methods used in Ref.$et al.$ \cite{collaborativesimilarity} by log-binning the degree-dependent collaborative similarity distributions to see more clear overall trends ranging from small-degree to high-degree. The collaborative similarity of developers in watching layer is negatively correlated with degree while, in contrast, a positive correlation is observed in the forking layer. This result agrees with the different usage of watching and forking. When a developer watches a project, he intends to just keep informed of some new trends and watching is used to broaden his interests. But when a developer forks a project, he intends to keep working on it and forking is used to keep more focused on a certain field. Thus, the collaborative similarity decreases while developers watch more projects and increases when developers forks more projects. The collaborative similarity of projects in watching layer is negatively correlated with degree while in forking layer no obvious correlations can be observed and the collaborative similarity fluctuates near the overall average collaborative similarity. The result can also be explained by the different usage of watching and forking functionalities. When a project becomes popular, it begins to attract developers from various fields to watch while developers who fork it always intend to keep working on it and are often from the same field.

	\section{Conclusion and discussion}

        We have empirically analyzed the multiple relations between developers and project on GitHub from the perspective of multilayer bipartite network. The degree distributions, the nearest neighbors' degree distributions and the collaborative similarity distributions are reported. Our results show that all degree distributions have a power-law form, especially, the degree distribution of projects in watching layer has double power-law form. We find a disassortative mixing pattern of developers and projects, that is, the nearest neighbors' degree and degree are negatively correlated. To study the diversity of interests, we apply the collaborative similarity index \cite{collaborativesimilarity} on our empirical analysis. The result shows that the collaborative similarity of both developers and projects negatively correlates with degree in watching layer, while a positive correlations is observed for developers in forking layer and no obvious correlation is observed for projects in forking layer. The reason behind this result is the different usage of these social functionalities provided by GitHub.

        Although a large number of researches have contributed to the understanding of bipartite networks and an increasing amount of researches are going to multiplex networks, combining multiple relations in bipartite networks from a systematic view and modeling them as multilayer bipartite networks seldom show up. Our study opens up a new perspective to study the multiple relations in bipartite networks.

    \section*{Acknowledgment}
        This work was funded by the National Natural Science Foundation of China (Grant Nos. 11275186, 91024026, and FOM2014OF001). We thank Ming Li for useful suggestions on the analysis methods.

    \section*{References}
    \bibliography{diversity-of-interests}

\end{document}